\documentclass[aps,showpacs,preprintnumbers,twocolumn,amsmath,amssymb,prl,floatfix]{revtex4}

\usepackage{graphicx}

\usepackage{psfrag}

\usepackage{ifpdf}

\ifpdf
\usepackage{epstopdf}
\usepackage[pdftex]{hyperref}
\else
\usepackage[hypertex,ps2pdf,dvips,colorlinks,urlcolor=blue,citecolor=green,linkcolor=red]{hyperref}
\fi
\begin{document}

\title{Exact three-body local correlations for excited
  states of the 1D Bose gas }
\author{M\'arton Kormos, Yang-Zhi Chou, and Adilet Imambekov}
\affiliation{Department of Physics and Astronomy, Rice
University, Houston, Texas 77005, USA}
\date{\today}

\def\be{\begin{equation}}
\def\ee{\end{equation}}
\def\bea{\begin{eqnarray}}
\def\eea{\end{eqnarray}}
\newcommand{\expct}[1]{\left\langle #1 \right\rangle}
\newcommand{\expcts}[1]{\langle #1 \rangle}
\newcommand{\ud}          {\mathrm d}
\newcommand\eps           {\varepsilon}
\newcommand\fii           {\varphi}
\newcommand\mc            {\mathcal}
\newcommand\LL            {Lieb--Liniger }
\newcommand\p             {\partial}
\newcommand\psid          {\psi^{\dagger}}
\renewcommand\th          {\theta}
\newcommand\kb            {k_\text{B}}

\begin{abstract}

We derive an exact analytic expression for the three-body local
correlations in the Lieb--Liniger model of 1D Bose gas with contact
repulsion.  The local three-body correlations control the
thermalization and particle loss rates in the presence of terms which
break integrability, as is realized in the case of 1D ultracold
bosons. Our result is valid not only at finite temperature but also
for a large class of non-thermal excited states in the thermodynamic
limit. We present finite temperature calculations in the presence of
external harmonic confinement within local density approximation, and
for a highly excited state that resembles an experimentally realized
configuration.

\end{abstract}

\pacs{67.85.-d,03.75.Hh,02.30.Ik,68.65.-k}

\maketitle

When ultracold bosons are confined to move in only one dimension (1D),
they provide a very clean realization~\cite{Olshanii98, Dweiss,
  cradle} of a seminal exactly solvable model introduced by Lieb and
Liniger (LL)~\cite{LL}.  Being an integrable model, it has very
special dynamics showing almost no relaxation in
experiments~\cite{cradle}. 

This fact stimulated lots of theoretical interest to understand the
thermalization of isolated 1D systems and the role of integrability as
well as its breaking in this
process~\cite{Rigol_GGE,Rigol_weak_nonintegrability,Rigol_Nature,Mazets_g3,thermal}.
In particular, it has been shown~\cite{Shlyapnikov,Mazets_g3}, that
virtual excitations of bosons to higher transverse modes of a
confining potential result in a weak three-body {\it local}
interaction that violates integrability of the many-body problem. Thus
it is important to understand three-body local correlations in the
absence of integrability breaking terms first.  Such correlations have
also been measured recently using analysis of particle
losses~\cite{3decay,haller}, density fluctuation
statistics~\cite{Bouchoule}, time-of-flight correlation
statistics~\cite{hodgman}, and scanning electron
microscopy~\cite{electron_microscopy3D}. They provide a very sensitive
test of coherence, and e.g., for Bose--Einstein condensates they
increase by a factor of $6=3!$ if the temperature is raised to be much
larger than the condensation
temperature~\cite{g3_theory3D,g3_experiment3D}.  In spite of the LL
model being integrable, analytical calculation of its correlation
functions is notoriously hard~\cite{KBI}.  
Two-body local correlations in equilibrium can be simply obtained
using the Hellmann--Feynman theorem and exact thermodynamics, and show
excellent agreement with experiments \cite{g2_socal}. Three body
local correlations were analytically calculated only at zero
temperature in a remarkable tour de force~\cite{g3}, as well as
numerically in Ref.~\cite{numerics}.

In this Letter, we exactly evaluate three-body local correlations in
the thermodynamic limit for a large class of excited states which can
be described by density matrices diagonal in the energy
representation. In particular, we apply our method at finite
temperatures and for highly excited states similar to the ones created
in experiments~\cite{cradle}, and we take into account external
harmonic confinement within local density approximation (LDA).
We note that local two-body correlations in 1D play the role of the
``contact" introduced by S. Tan~\cite{Tan, Tan1D}. Similarly,
three-body local correlations correspond to a three-body contact which
is being actively explored~\cite{3p_contact}. Our exact results
provide an important benchmark for such theories, as well as for
numerical methods for simulating field theories in 1D~\cite{MPS}.

\paragraph*{The model.---} The LL model describes a system of identical
bosons in 1D interacting via a Dirac-delta potential.
The Hamiltonian in second quantized formulation is given by 
\be
H= \int_0^L\ud x\,\frac{\hbar^2}{2m}\left(\p_x\psid\p_x\psi+c\,\psid\psid\psi\psi\right)\,,
\label{eq:H_NLS}
\ee
where $c>0$ in the repulsive regime we wish to study, and $m$ is the
atomic mass. The dimensionless coupling constant is given by
$\gamma=c/n$, where $n=N/L$ is the density of the gas. We will express
temperature $T$ in dimensionless units $\tau=T/T_{\text{D}}$, where
$T_{\text{D}}=\hbar^2 n^2/(2m\kb)$ is the quantum degeneracy
temperature.

The exact thermodynamics of the model can be obtained via Bethe Ansatz
\cite{LL,KBI}.  Each eigenstate of the system with $N$ particles on a
ring of circumference $L$ is characterized by a distinct set of
quantum numbers $\{I_j\}$ that are integers (half-integers) for $N$
odd (even). The wave function can be expressed in terms of $N$
quasimomenta $\{p_j\}$ that satisfy a set of algebraic equations
\be
Lp_j+\sum_{k=1}^N\theta(p_j-p_k) = 2\pi I_j\,,
\label{eq:BAE}
\ee
where $\theta(p)=2\arctan(p/c)$. The wave function is identically
zero if any two of the $\{I_j\}$ coincide, which is reminiscent
of the Pauli principle for fermions.  In the Tonks--Girardeau (TG)
limit $c\to \infty$, $\{I_j\}$ correspond to the quantum numbers of
occupied single-particle states of free fermions.
%

In the thermodynamic limit, if one wants to consider a mixed state
diagonal in the energy basis, this is achieved by introducing a
filling fraction $0<f_I<1$ in the space of quantum numbers, which
plays a role similar to the occupation number of free fermions.
All results of the present Letter are valid for $f_I$ which have a
finite thermodynamic limit at constant $I/N$; the limiting function
should be piecewise continuous and normalized. For calculations, it is
more convenient to define a function $f(p)$ in terms of the
quasimomenta: denoting by $\rho(p)$ the maximal allowed density of
quasimomenta in the vicinity of $p$, the quasimomenta density for a
mixed state is given by $f(p)\rho(p)$.

Since all quasimomenta are coupled to each other by
Eq.~\eqref{eq:BAE}, the density $\rho(p)$ is not independent of
$f(p)$: it satisfies the integral equation and normalization condition
\begin{subequations}
\label{eq:rho}
\begin{align}
\rho(p)&=\frac1{2\pi} + \int\frac{\ud p'}{2\pi}\,f(p')\,\fii(p-p')\,\rho(p')\,,\label{eq:rhoeq} \\
n&=\int\ud p \,f(p)\,\rho(p)\,,\label{eq:rhonorm}
\end{align}
\end{subequations}
with the kernel $\fii(p)=2c/(p^2+c^2)$. In thermal equilibrium, $f(p)$
has to satisfy a set of nonlinear integral
equations~\cite{yang,epaps}, but our results will be valid for more
general $f(p)$.

\paragraph*{Local correlations.---} The local $k$-body correlation
functions are defined as  
\be
g_k(\gamma,\tau)=\frac{\expct{\psi^{\dagger k}(x)\psi^k(x)}}{n^k}\,.
\ee
The first two of them are relatively easy to calculate: $g_1=1$,
while $g_2$ in equilibrium is given by the Hellmann--Feynman theorem
\cite{g2_socal, epaps}.

Here we report the results for $k=2$ and $k=3$ for general $f(p)$,
which can be written in terms of functions $h_m(p)$ ($m=1,2$)
satisfying the following integral equations:
\be
h_m(p)=p^m+\int\frac{\ud p'}{2\pi}\,f(p')\fii(p-p')h_m(p')\,.
\label{eq:h}
\ee 
In the case of $g_2$ the final formula is
\be
g_2(\gamma,\tau) = \frac{2\gamma^2}{c^3} \int \frac{\ud p}{2\pi}\,f(p)
\left[2\pi\rho(p)\,p^2 - h_1(p)\,p\right]\,,
\label{eq:g2gen}
\ee
which agrees with the result of the Hellmann--Feynman theorem for
thermal equilibrium \cite{epaps}, but is more general.
Similarly, for $k=3$ the final expression is given by
\begin{multline}
g_3(\gamma,\tau) =
\frac{\gamma^3}{c^5} \int\frac{\ud p}{2\pi}\,f(p)\bigg[(p^4+c^2p^2)2\pi \rho(p)-\\
\left(4p^2+(1+2/\gamma)c^2\right)p\,h_1(p)+3p^2 h_2(p)\bigg]+\\
\frac{2\gamma^3}{c^4}\left(\int\ud p\,f(p)\rho(p)p\right)^2\,.
\label{eq:result}
\end{multline}
In the case when $f(p)$ is even, in equilibrium for example, the last term in
Eq.~\eqref{eq:result} is zero because the integrand is odd in
$p$. Both Eq.~\eqref{eq:g2gen} and Eq.~\eqref{eq:result} are Galilean
invariant expressions \cite{epaps}. 
In the following we will first consider an equilibrium case and then will
proceed to highly excited states.

\begin{figure}[t]
\centerline{
\scalebox{0.3}{\includegraphics{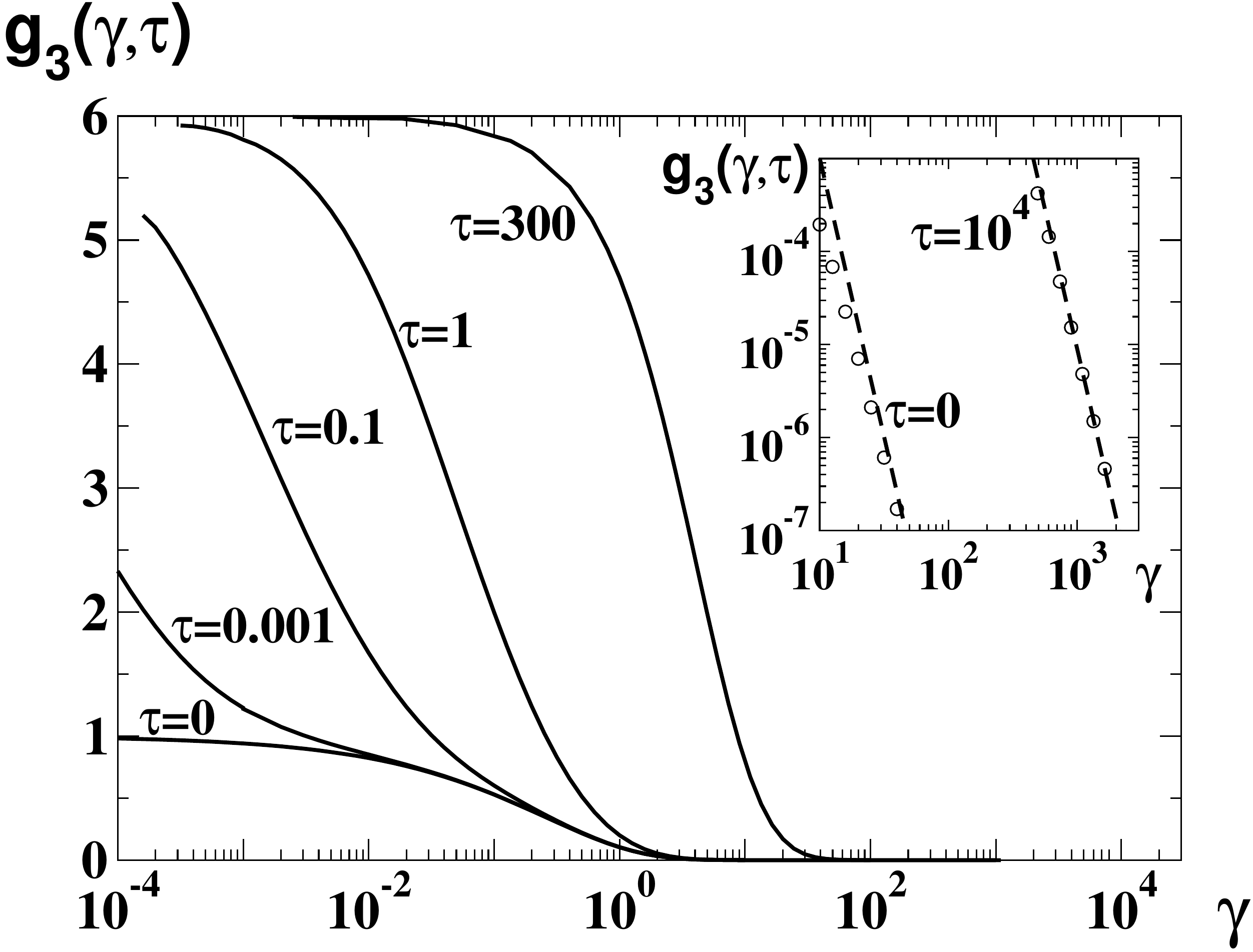}}}
\caption{Local three-body correlator $g_3(\gamma,\tau)$ as a function
  of dimensionless coupling $\gamma=c/n$ for the uniform system at
  fixed dimensionless temperature $\tau=T/T_{\text{D}}$, where
  $T_{\text{D}}=\hbar^2n^2/(2m\kb)$ is the quantum degeneracy
  temperature. The inset shows the large $\gamma$ asymptotic
    behavior on a log-log scale.}
\label{fig:fig1}
\end{figure}


In Fig.~\ref{fig:fig1} we plot the result in thermal equilibrium for
fixed $\tau$ as a function of the coupling $\gamma$. In particular, at
zero temperature our result agrees with that of Ref.~\cite{g3}, up to
the precision of the numerical evaluation of both expressions,
$\approx10^{-3}$.  The behavior of $g_3$ is qualitatively similar to
that of $g_2$ analyzed in Ref.~\cite{g2_socal} and it distinguishes
three different physical regimes: (a)
$\gamma\gtrsim\text{max}(1,\sqrt\tau)$, strong coupling (TG) regime,
$g_3 \ll 1$; (b) $\tau^2\lesssim\gamma\lesssim1$, quasicondensate
regime, $g_3 \approx 1$; (c)
$\gamma\lesssim\text{min}(\tau^2,\sqrt\tau)$, decoherent regime, $g_3
\approx 6$.
%
%
In the inset of Fig.~\ref{fig:fig1} the large $\gamma$ asymptotics
are plotted together with the analytic forms of Ref.~\cite{g2_socal}:
$g_3\sim16\pi^6/(15\gamma^6)$ for $\tau=0$ and
$g_3\sim9\tau^3/\gamma^6$ for $\gamma^2\gg\tau\gg1$.

\begin{figure}[t]
\centerline{
\scalebox{0.3}{\includegraphics{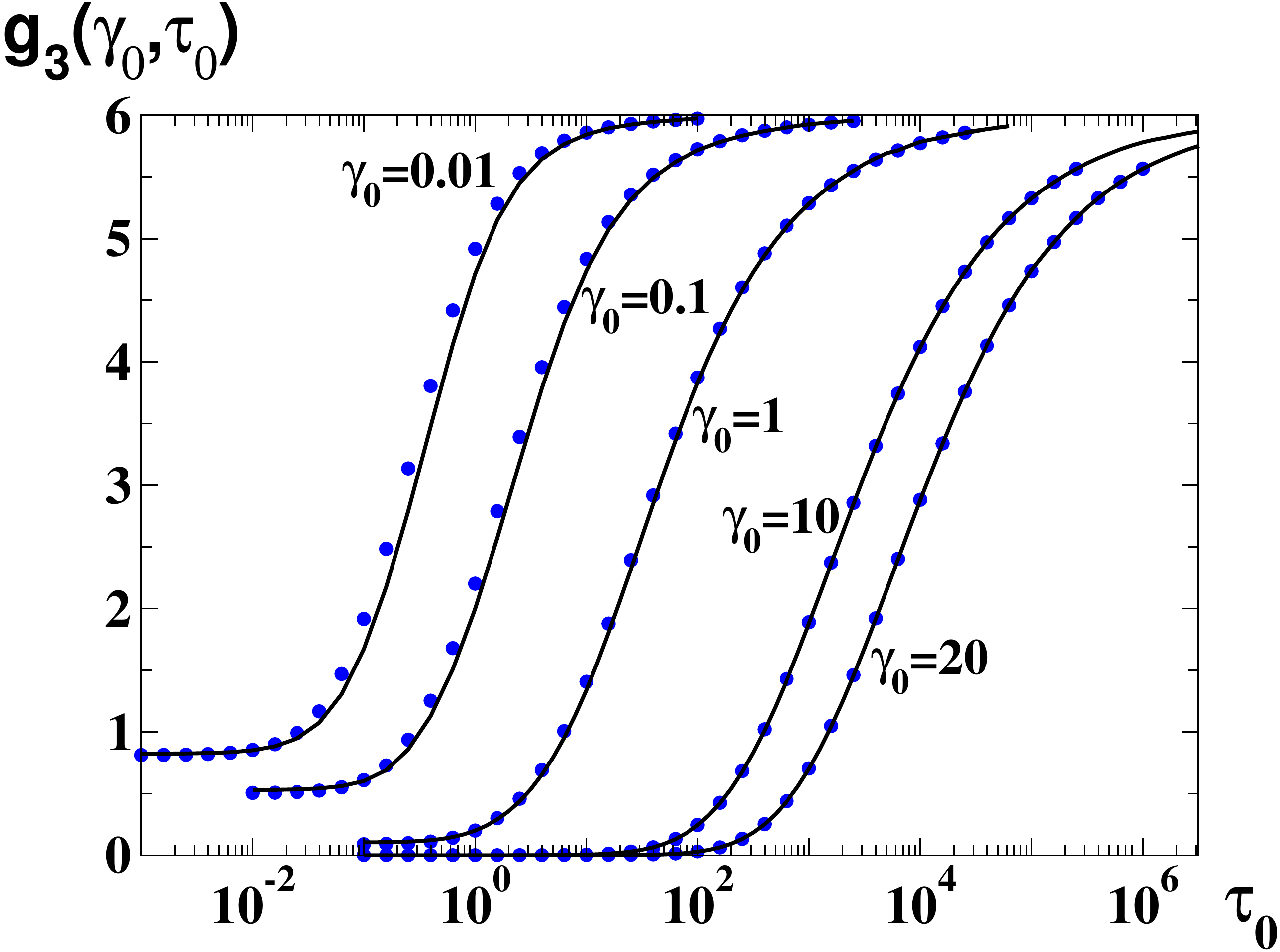}}}
\caption{Local three-body correlators
  $\overline{g_3}(\gamma_0,\tau_0)$ averaged over the trap (dots) and
  $g_3(\gamma_0,\tau_0)$ in the center (solid lines), as a function of
  the dimensionless temperature $\tau_0$ for fixed dimensionless
  coupling $\gamma_0$ in the center of the trap. At high temperature,
  $g_3$ approaches the value $6=3!$, reflecting Bose statistics.}
\label{fig:fig2}
\end{figure}

\paragraph*{Harmonic traps.---} 
Next we turn to the experimentally more realistic
case of atoms confined in a waveguide with a harmonic longitudinal
potential. The 1D regime is reached if $\mu,\kb
T\ll\hbar\omega_\perp$, where $\mu$ is the chemical potential and
$\omega_\perp$ is the transverse oscillator frequency \cite{petrov}.
If the density profile in the trap varies smoothly,
the correlations can be calculated by combining our exact results with
LDA~\cite{LDA}.
The relevant properties of the gas can be characterized by the LL
coupling $\gamma_0$ and the temperature parameter $\tau_0$ at the
center of the trap.
In Fig.~\ref{fig:fig2} we plot the three-body correlator
$g_3(\gamma_0,\tau_0)$ at the trap center and the normalized average,
$\overline{g_3}(\gamma_0,\tau_0)=\int\ud x \expct{\psi^{\dagger 3} (x)
  \psi^3(x)} /\left(\int\ud x\,n^3(x)\right)$, against the
dimensionless temperature $\tau_0$ for different fixed values of
$\gamma_0$.  Similarly to the results of Ref.~\cite{LDA} for the case
of $g_2$,
we find that unless the coupling $\gamma_0$ is very small,
$g_3(\gamma_0,\tau_0)\approx \overline{g_3}(\gamma_0,\tau_0)$
at any temperature. 

The curves in Fig.~\ref{fig:fig2} are related to
the observed change in time of the particle loss rate in
Ref.~\cite{haller}. With increasing temperature, the three-body
correlations grow according to our result, which leads to a higher
probability of inelastic three-particle processes which in turn raises
further the temperature of the gas. This positive feedback causes a
non-trivial dependence of the particle loss on time, and a detailed
analysis of heating mechanisms is needed to describe the time
dependences of the loss rates.

\paragraph*{Three-body correlations for highly excited states.---}

\begin{figure}[t]
\centerline{
\scalebox{0.3}{\includegraphics{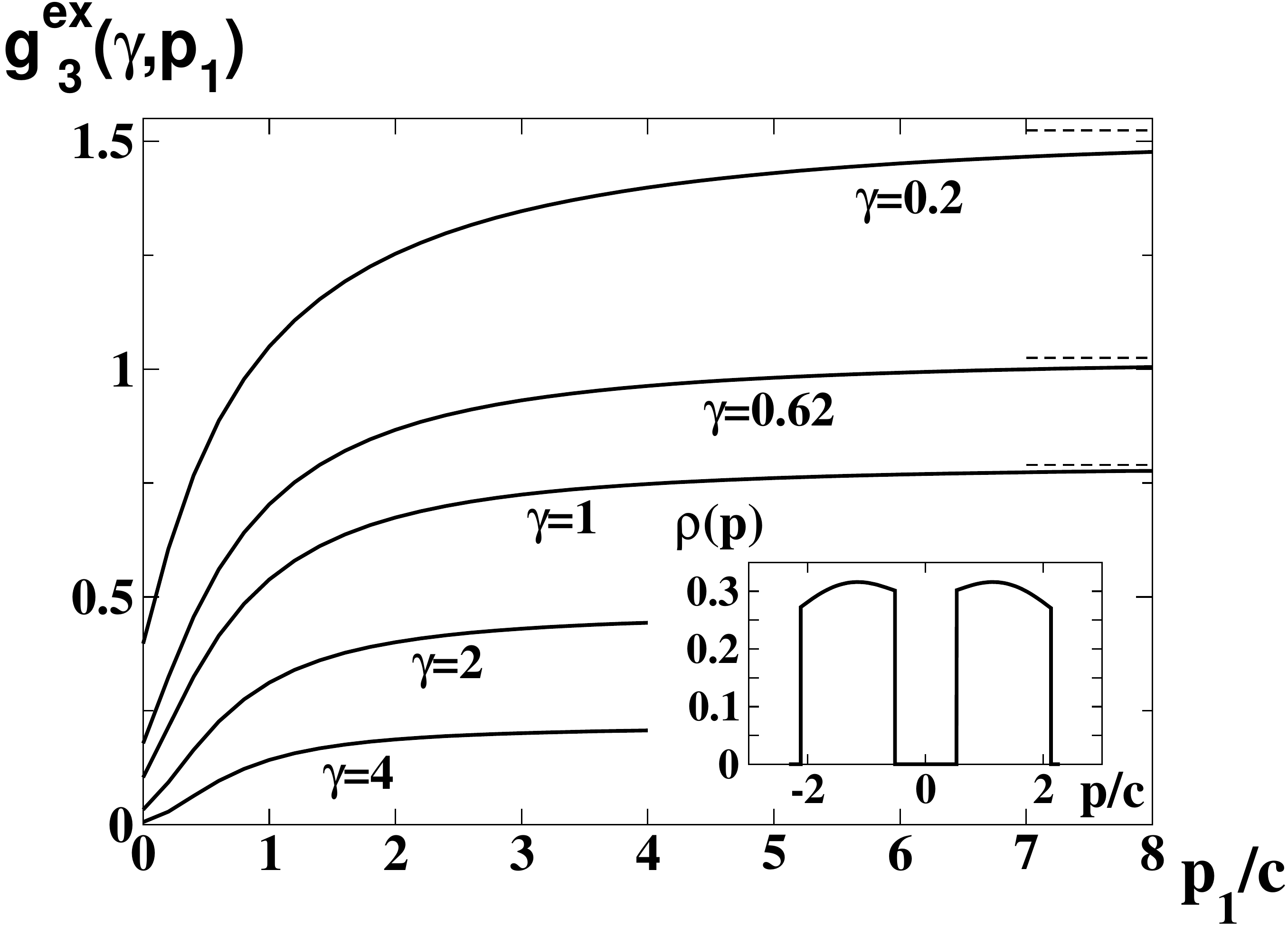}}}
\caption{Local three-body correlator $g^{\text{ex}}_3(\gamma,p_1)$ as
  a function of $p_1$ for different values of $\gamma$. The excited
  state is characterized by the inner ``Fermi quasimomentum''
  $p_1$. The horizontal lines correspond to $g^{ex}_3 (\gamma,
  \infty)=\left[g_3(2\gamma)+9g_2(2\gamma)\right]/4$. The inset
  illustrates a typical quasimomentum density for $\gamma=1$,
  $p_1=0.5c$.}
\label{fig:fig3}
\end{figure}

Since Eq.~\eqref{EQ:GK} is valid for general distributions $f(p)$
\cite{Pozsgay}, we can use our results
\eqref{eq:g2gen},\eqref{eq:result} in situations where the system is
neither in equilibrium nor is in its ground state.  We will illustrate
this by considering a state which is motivated by the experiment of
Kinoshita \textit{et al.} \cite{cradle}, where each atom was put in a
momentum superposition state, after which the two clouds performed
many oscillations without observable thermalization. The state created
in the experiment is not an eigenstate, and the harmonic trap might
play an important role. However, let us consider here a simple
``caricature'' eigenstate which might capture the behavior of $g_3$,
and hence the role of integrability breaking, in these experiments.
This state is characterized by an $f(p)$ consisting of two disjoint
rectangular ``Fermi steps" symmetric with respect to $p=0$ at zero
temperature: $f(p)=\theta(p^2-p_1^2)-\theta(p^2-p_2^2)$ with
$p_2>p_1>0$ (see inset of Fig.~\ref{fig:fig3}).  We are interested in
the dependence of $g^{\text{ex}}_3(\gamma,p_1)$ on the ``inner Fermi
quasimomentum'' $p_1$.  If $p_1$ is fixed then the ``outer Fermi
quasimomentum'', $p_2$, is determined from the normalization
\eqref{eq:rhonorm}. The momentum kick in Ref.~\cite{cradle}
corresponds to $p_1/c$ of order one.  In Fig.~\ref{fig:fig3} we plot
$g^{\text{ex}}_3(\gamma,p_1)$ for fixed values of $\gamma$ as a
function of $p_1$. We find that the correlations grow with the
momentum of the kick and they can become greater than 1. For large
$p_1$, the quasimomentum distributions of left and right goers become
approximately independent of each other. However, to obtain the
correct limit as $p_1 \to \infty$ one needs to take into account
deviations of $\theta(2p_1/c\gg 1)$ in Eq.~\eqref{eq:rho} from
$\pi$. This results in $g^{ex}_3 (\gamma,
\infty)=\left[g_3(2\gamma)+9g_2(2\gamma)\right]/4$; in particular,
$g^{ex}_3 (\gamma \to 0, \infty)=5/2$. Similarly, $g^{ex}_2 (\gamma,
\infty)=\left[g_2(2\gamma)+2\right]/2$, and $g^{ex}_2 (\gamma \to 0,
\infty)=3/2$~\cite{epaps}.

\paragraph*{Derivation of Eqs.~\eqref{eq:g2gen} and \eqref{eq:result}.---}
In Ref.~\cite{Kormos} a novel method was proposed to calculate the $g_k$
correlators based on the observation that the LL model can be
viewed as the combined non-relativistic, weak coupling limit of the
sinh--Gordon model. The resulting formula reads as \cite{Kormos,Pozsgay}
\be
g_k = 
\sum_{s=k}^{\infty}\frac1{s!}\int \prod_{j=1}^s\frac{\ud
p_j}{2\pi}\,f(p_j)\,\, \gamma^kF^{(k)}_s(p_1,\dots,p_s)\,,
\label{EQ:GK}
\ee
where the form factors $F^{(k)}_s(p_1,\dots,p_s)$ are the infinite
volume $s$-particle diagonal matrix elements of the operator
$\psi^{\dagger k}\psi^k$, which can be obtained from known
sinh--Gordon form factors~\cite{KMP,giuseppe,babkar}.  These series
were investigated previously by truncating them after the first few
terms~\cite{Kormos}. Here we resum these series to all
orders obtaining closed analytical expressions for the local
correlations.

It has been proven in Ref.~\cite{Pozsgay} that 
\begin{subequations}
\begin{alignat}{2}
F^{(1)}_s&=\frac1c &\sum_P& \,\fii(p_{12}) \fii(p_{23})\dots \fii(p_{s-1,s}) \,,\\
F^{(2)}_s&=\frac1{c^3} &\sum_P& \,\fii(p_{12}) \fii(p_{23})\dots \fii(p_{s-1,s})\,\,p_{1,s}^2 \,,
\end{alignat}
and based on evaluations performed in Mathematica for the first few
$F^{(3)}_s$ we conjecture
\begin{multline}
F^{(3)}_s=\frac1{c^5}\sum_P \,\fii(p_{12})\fii(p_{23})\dots
\fii(p_{s-1,s})\times\\
\frac12\,p_{1,s}\left[p_{1,s}^3-(p_{12}^3+p_{23}^3\dots+p_{s-1,s}^3)\right]\,,\label{eq:FF3}
\end{multline}
\end{subequations}
where $p_{ij}=p_i-p_j$ and $\sum_P$ denotes a sum over all
permutations of $\{p_j\}$.
Below we will illustrate how series \eqref{EQ:GK} can be analytically
resummed for $g_2$, and details of similar calculations for $g_3$ are
presented in EPAPS~\cite{epaps}.

We will use abbreviations $\tilde\ud p=\ud p/(2\pi)f(p)$ and
$\fii_{ij}=\fii(p_{ij})$. Using the symmetries of the integrand in
Eq.~\eqref{EQ:GK}, we have for $c^3g_2(\gamma,\tau)/(2\gamma^2)$
\begin{multline}
\label{eq:deriv}
\frac12\sum_{s=2}^{\infty}
\int\tilde\ud p_1\dots 
\int\tilde\ud p_s\,
\fii_{12}\dots \fii_{s-1,s}(p_1-p_s)^2=  \\
\int\tilde\ud p_1\,p_1^2
\left[ 
\int\tilde\ud p_2\,
\fii_{12} +
\int\tilde\ud p_2\!\!
\int\tilde\ud p_3\,
\fii_{12}\fii_{23}+\dots\right]-\\
\int\!\tilde\ud p_1p_1\!\!
\left[ 
\int\!\tilde\ud p_1\, \fii_{12}\,p_2 +
\int\!\tilde\ud p_2\!\!
\int\!\tilde\ud p_3\,
 \fii_{12}\fii_{23}\,p_3+\dots\right]=\\
\int\tilde\ud p_1\,p_1^2 \left[2\pi
\rho(p_1)-1\right] - 
\int\tilde\ud p_1\,p_1 
\left[h_1(p_1)-p_1\right],
\end{multline}
where the terms in the first square bracket coincide with
the iterative solution of the integral equation
\eqref{eq:rhoeq}. Similarly, comparison of the terms in the second
bracket with the iterative solution of Eq.~\eqref{eq:h} leads to
$h_1(p_1) - p_1$. Now the second terms in the parentheses cancel each
other and we obtain Eq.~\eqref{eq:g2gen}.

In summary, we derived an {\em exact} formula for the local three-body
correlation in a paradigmatic system, the 1D Lieb--Liniger Bose
gas. Given that exact expressions for correlation functions are scarce
even in integrable models, we emphasize the analytic nature of our
result. Our non-perturbative formula is valid at any temperature for
arbitrary value of the coupling $\gamma$, and for a large class of
excited states of the system. The result can open the window to an
analytic treatment of integrability breaking perturbations and
thermalization in nearly integrable systems.

\begin{acknowledgments}
We acknowledge funding from The Welch Foundation, Grant No.\ C-1739, from
the Sloan Foundation and from the NSF Career Award No.\ DMR-1049082.
\end{acknowledgments}

\paragraph{Note added.---} While this paper was under review, we learned about
Ref.~\cite{Pozsgay_new} were some of our results
were rederived and generalized.

\clearpage
\newpage

\clearpage 

\onecolumngrid

\begin{center}
{\Large Supplementary Material for EPAPS \\ 
Exact three-body local correlations for excited
  states of the 1D Bose gas }
\end{center}

\section{Thermal equilibrium}

In thermal equilibrium \cite{yang}, the filling fraction can be written as 
$f(p)=(1+e^{\eps(p)})^{-1}$, where the pseudo-energy $\eps(p)$
satisfies the nonlinear integral equation
\be
\eps(p) = -\frac{\mu}{\kb T}+\frac{\hbar^2p^2}{2m\kb T}-
\int\frac{\ud p'}{2\pi}\, \fii(p-p')\log\left(1+e^{-\eps(p')}\right)\,,
\label{eq:eps}
\ee
and the free energy is given by
\be
F=L\,\left(\mu\, n-\kb T\int
\frac{\ud p}{2\pi}\,\log(1+e^{-\eps(p)})\right)\,.
\label{eq:FreeE}
\ee
Eq.~\eqref{eq:eps} is coupled to the equations for the density
\begin{subequations}
\label{eq:rho2}
\begin{align}
\rho(p)&=\frac1{2\pi} + \int\frac{\ud p'}{2\pi}\,f(p')\,\fii(p-p')\,\rho(p')\,,\\
n&=\int\ud p \,f(p)\,\rho(p)\,,
\end{align}
\end{subequations}
At $T=0$, Eqs.~\eqref{eq:rho2} and \eqref{eq:eps} decouple and
$1/(1+e^{\eps(p)})$ becomes a Fermi step function
$f(p)=\theta(p^2-p_\text{F}^2)$ where $p_\text{F}$ is determined from
$\int_{-p_\text{F}}^{p_\text{F}}\ud p\rho(p)=n$.

\section{Dimensionless variables}

It is useful to introduce dimensionless quasimomenta $q=p/c$, 
and to change slightly the notation for the densities: $\rho(c\,q)\to \rho(q)$, and similarly for $f(q)$ and $\eps(q)$. Then 
the equations for the density become
\begin{subequations}
\label{eq:rhoq}
\begin{align}
\rho(q)&=\frac1{2\pi} + \int\frac{\ud q'}{2\pi}\,f(q')\,\fii(q-q')\,\rho(q')\,,\label{eq:rhoeqq} \\
\frac1\gamma&=\int\ud q \,f(q)\,\rho(q)\,,\label{eq:rhonormq}
\end{align}
\end{subequations}
where the kernel is $\fii(q)=2/(q^2+1)$. In thermal equilibrium, $f(q)=(1+e^{\eps(q)})^{-1}$, 
and the equation for $\eps(q)$ is
\be
\eps(q) = -\alpha+\frac{q^2\gamma^2}{\tau}-
\int\frac{\ud q'}{2\pi}\, 
\fii(q-q')\log\left(1+e^{-\eps(q')}\right)\,,
\label{eq:epsq}
\ee
The dimensionless chemical potential $\alpha=\mu/\kb T$ needs to
be determined by the self-consistent solution of Eqs.\
\eqref{eq:rhoq},\eqref{eq:epsq}. Once $\eps(q)$ is found, the free energy
$F$ is given by
\be
F=L\,\tau \frac{\hbar^2n^3}{2m} \left(\alpha-\gamma\int
\frac{\ud q}{2\pi}\,\log(1+e^{-\eps(q)})\right)\,.
\label{eq:F}
\ee
The dimensionless form of the functions $h_m$ are defined as
\be
h_m(q)=q^m+\int\frac{\ud q'}{2\pi}\,f(q')\fii(q-q')h_m(q')\,.
\label{eq:hq}
\ee

\section{Hellmann--Feynman theorem}

For completeness, in this section we derive a formula
which was established earlier in Ref.~\cite{Kormos} based
on the Hellmann--Feynman theorem. This theorem states that
\be
\frac2L \expct{\frac{\partial H}{\partial g} } =\expct{\psid\psid\psi\psi}=2\frac{\ud}{\ud g}\left(\frac{
  F}L\right)\,,
\label{eq:HF2}
\ee
where the free energy can be calculated from Eq.~\eqref{eq:F}. 
In dimensionless variables,
\be
g_2(\gamma,\tau)=\tau\frac{\ud}{\ud\gamma}\left(\alpha-\gamma\int
\frac{\ud q}{2\pi}\,\log(1+e^{-\eps(q)})\right)\,.
\ee
Now we recast it in a form which does not exhibit
explicit derivatives. We will need the derivative of
Eq.~\eqref{eq:epsq} with respect to $\gamma$:
\be
\eps'(q) \equiv\frac{\ud \eps(q)}{\ud \gamma}=-\alpha'+\frac{2q^2\gamma}{\tau}+
\int\frac{\ud \bar{q}}{2\pi}\, 
\fii(q-\bar{q})\frac{\eps'(\bar{q})}{1+e^{\eps(\bar{q})}}\,.
\label{eq:depdg}
\ee
We start by writing
\be
g_2(\gamma,\tau)=\tau\alpha'-\tau\int
\frac{\ud q}{2\pi}\,\log(1+e^{-\eps(q)})+\tau\gamma \int
\frac{\ud q}{2\pi}\,\frac{\eps'(q)}{1+e^{\eps(q)}}\,,
\label{eq:step1}
\ee
where the prime denotes derivative with respect to $\gamma$. The key
step is substituting for $1/2\pi$ in the last integral the rest of 
Eq.~\eqref{eq:rhoeqq}:
\begin{multline}
\int \frac{\ud q}{2\pi}\,
\frac{\eps'(q)}{1+e^{\eps(q)}} =
\int \ud q\,
\left(\rho(q)-\int\frac{\ud \bar{q}}{2\pi}\,\fii(q-\bar{q})\,\frac{\rho(\bar{q})}{1+e^{\eps(\bar{q})}}\right)
\frac{\eps'(q)}{1+e^{\eps(q)}} = \\
\int \ud q\,\rho(q)\frac{\eps'(q)}{1+e^{\eps(q)}} -
\int\ud \bar{q}\frac{\rho(\bar{q})}{1+e^{\eps(\bar{q})}}
\int\frac{\ud q}{2\pi}\,
\fii(\bar{q}-q)\frac{\eps'(q)}{1+e^{\eps(q)}}\,.
\end{multline}
In the last step we used the fact that the kernel is an even
function. Now we use Eq.~\eqref{eq:depdg} to express the convolution
term and continue with the equalities as
\begin{align}
\int \frac{\ud q}{2\pi}\,
\frac{\eps'(q)}{1+e^{\eps(q)}} &= 
\int \ud q\,\rho(q)\frac{\eps'(q)}{1+e^{\eps(q)}} -
\int\ud \bar{q}\frac{\rho(\bar{q})}{1+e^{\eps(\bar{q})}}
\left(\eps'(\bar{q})+\alpha'-\frac{2\bar{q}^2\gamma}{\tau}\right) = \nonumber\\
&-\frac{\alpha'}{\gamma} + \frac{2\gamma}{\tau}
\int\ud \bar{q}\frac{\rho(\bar{q})}{1+e^{\eps(\bar{q})}}\bar{q}^2\,,
\end{align}
where the terms with $\eps'$ dropped out and we used
Eq. \eqref{eq:rhonormq}. Plugging this result into Eq.~\eqref{eq:step1} even
the $\alpha'$ terms cancel and we finally arrive at
\be
g_2(\gamma,\tau)=2\gamma^2
\int\ud q\frac{\rho(q)}{1+e^{\eps(q)}}q^2  -\tau\int
\frac{\ud q}{2\pi}\,\log(1+e^{-\eps(q)})\,.
\ee
We can rewrite the second integral using partial integration:
\be
g_2(\gamma,\tau)=2\gamma^2
\int\ud q\frac{\rho(q)}{1+e^{\eps(q)}}q^2  -\tau\int
\frac{\ud q}{2\pi}\,q\,\frac1{1+e^{\eps(q)}}\frac{\ud\eps(q)}{\ud q}\,.
\label{eq:g2alt}
\ee
Differentiating Eq.~\eqref{eq:epsq} with respect to $q$ we obtain 
\be
\frac{\ud \eps(q)}{\ud q}=\frac{2\gamma^2}{\tau}\,q+
\int\frac{\ud \bar{q}}{2\pi}\,\frac1{1+e^{\eps(\bar q)}} 
\fii(q-\bar{q})\frac{\ud\eps(\bar q)}{\ud \bar q}\,,
\ee
which shows that $\ud\eps(q)/\ud q$ satisfies
the same integral equation as $2\gamma^2/\tau\,h_1(q)$ (see
Eq.~\eqref{eq:hq}). Thus expression \eqref{eq:g2alt} is clearly the special
case of the general result 
\be
g_2(\gamma,\tau)=2\gamma^2
\int\ud q\,f(q)\rho(q)\,q^2  - 2\gamma^2\int
\frac{\ud q}{2\pi}\,f(q)h_1(q)q\,,
\label{eq:g2genq}
\ee
for equilibrium when $f(q)=1/(1+e^{\eps(q)})$.

In the original dimensionful variables one also has to take care of
the coupling constanst dependence of the kernel. It is interesting to
note that the analogous derivation for the $T=0$ case is more subtle
because then one has to differentiate the boundary of integration
explicitly.

\section{Summing up the series \texorpdfstring{\eqref{EQ:GK} for
    $k=3$}{for k=3}}

Every permutation in the form factors \eqref{eq:FF3} contains all the
momenta and not only the first and the last few of them, which seems
to render the procedure used in the main text for $k=2$ unfeasible. However, this
difficulty can be overcome by rewriting the form factors exploiting
their special structure:
\begin{multline}
F^{(3)}_s=\frac12\sum_P \fii(q_{12})\fii(q_{23})\dots \fii(q_{s-1,s}) \left[q_{1,s}^4-q_{1,s}\sum_{d=1}^{s-1}q_{d,d+1}\left(\frac2{\fii(q_{d,d+1})}-1\right)\right]=\\
\frac12\sum_P \left\{
\fii(q_{12})\fii(q_{23})\dots \fii(q_{s-1,s})\left(q_{1,s}^4+q_{1,s}^2\right)-2q_{1,s}\sum_{d=1}^{s-1}\fii(q_{12})\dots \fii(q_{d-1,d})\fii(q_{d+1,d+2}) \dots \fii(q_{s-1,s})q_{d,d+1}\right\}\,,
\label{eq:FF}
\end{multline}
where in the last term the kernel $\fii(q_{d,d+1})$ is
missing. We used here the explicit form of the kernel and
$\sum_{d=1}^{s-1}q_{d,d+1}=q_{1,s}$. In the last term in the second
line the chain of kernels is broken and the polynomial only depends on
momenta at the ends of the resulting chains, which allows us to use the
same technique as before.

\subsection{First term}

Let us start with expanding the polynomial part
\begin{align}
q_{1,s}^4+q_{1,s}^2 = &(q_1^4+q_1^2) -(4q_1^3+q_1)q_s +3q_1^2q_s^2+\\
&(q_s^4+q_s^2) -(4q_s^3+q_s)q_1 +3q_s^2q_1^2\,.
\end{align}
The second line of the expansion can be obtained from the first line
by the transformation $q_1\leftrightarrow q_s$, that is, by a
reflection of the indices. Since the kernel $\fii(q)$ is an even
function, the form factor is invariant under this transformation
(actually, every single term in the permutation sum is invariant). We
are thus allowed to focus on the terms in the first line, taking the
second line into account by multiplying the result by 2 which cancels
the overall $1/2$.

\begin{itemize}
\item Terms containing $q_1$ only.\\
The subseries corresponding to the terms $(q_1^4+q_1^2)$ is
\begin{multline}
\int\tilde\ud q_1\, (q_1^4+q_1^2)
\left(\int\tilde\ud q_2\int\tilde\ud q_3\,
\fii(q_{12})\fii(q_{23}) +
\int\tilde\ud q_2
\int\tilde\ud q_3
\int\tilde\ud q_4\,
\fii(q_{12})\fii(q_{23})\fii(q_{34})+\dots\right)=\\
\int\tilde\ud q_1(q_1^4+q_1^2)\left[2\pi \rho(q_1)-1-\int\tilde\ud q_2\, 
\fii(q_{12})\right]\,,
\label{eq:1}
\end{multline}
where we used Eq.~\eqref{eq:rhoeqq}.
\item Terms containing $q_1,q_s$.\\
For the term $-(4q_1^3+q_1)q_s$ we have the series
\begin{multline}
-\int\tilde\ud q_1\, (4q_1^3+q_1)
\left(\int\tilde\ud q_2\int\tilde\ud q_3\,
\fii(q_{12})\fii(q_{23})q_3 +
\int\tilde\ud q_2
\int\tilde\ud q_3
\int\tilde\ud q_4\,
\fii(q_{12})\fii(q_{23})\fii(q_{34})q_4+\dots\right)=\\
-\int\tilde\ud q_1(4q_1^3+q_1)
\left[h_1(q_1)-q_1-\int\tilde\ud q_2\, 
\fii(q_{12})q_2\right]\,,
\label{eq:2}
\end{multline}
where in the last step we used Eq.~\eqref{eq:hq}. 
\item Finally, in the case of $3q_1^2q_s^2$ we are similarly led to
\begin{multline}
\int\tilde\ud q_1\, 3q_1^2
\left(\int\tilde\ud q_2\int\tilde\ud q_3\,
\fii(q_{12})\fii(q_{23})q_3^2 +
\int\tilde\ud q_2
\int\tilde\ud q_3
\int\tilde\ud q_4\,
\fii(q_{12})\fii(q_{23})\fii(q_{34})q_4^2+\dots\right)=\\
3\int\tilde\ud q_1q_1^2
\left[h_2(q_1)-q_1^2-\int\tilde\ud q_2\, 
\fii(q_{12})q_2^2\right]\,.
\label{eq:3}
\end{multline}
\end{itemize}

Combining Eqs.~(\ref{eq:1}-\ref{eq:3}) we find that several terms
cancel each other and we are left with
\begin{multline}
\int\tilde\ud q\,
(q^4+q^2)2\pi \rho(q)-
\int\tilde\ud q\, 
(4q^3+q)h_1(q)+
3\int\tilde\ud q\, 
q^2h_2(q)\\
-\int\tilde\ud q_1\int\tilde\ud q_2\,\fii(q_{12})\left((q_1^4-4q_1^3q_2+3q_1^2q_2^2)+(q_1^2-q_1q_2)\right)\,.
\end{multline}
Using the symmetry of $\fii(q_1-q_2)$ and then its explicit expression
we can write the last term in the nice form 
\be 
-\frac12\int\tilde\ud
q_1\int\tilde\ud q_2\,\fii(q_{12})(q_{12}^4+q_{12}^2) =
-\frac12\int\tilde\ud q_1\int\tilde\ud q_2\,2q_{12}^2\,, 
\ee
so the total contribution from the first term of the form factor is
\be
\int\tilde\ud q\,(q^4+q^2)2\pi \rho(q)-
\int\tilde\ud q\,(4q^3+q)h_1(q)+
3\int\tilde\ud q\,q^2h_2(q)
-\int\tilde\ud q_1\int\tilde\ud q_2\,q_{12}^2\,.
\label{eq:I}
\ee

\subsection{Second term}

The polynomial part is
\be
q_{1,s}q_{d,d+1}=(q_1q_d-q_1q_{d+1}) + (q_{d+1}q_s-q_d q_s)\,.
\label{eq:2ndterm}
\ee
Again, the second two terms give the same contribution as the first
two due to the sum over $d$, so we focus on the first two terms and
multiply the result by 2 in the end.  Due to the missing kernel, the
$s$-fold multiple integrals split into the product of a $d$-fold and
an $(s-d)$-fold integral.

\begin{itemize}
\item $-2q_1q_d$.\\
Taking into account all the prefactors, the subseries corresponding to
the first term in the first parenthesis is
\begin{multline}
-2\int\tilde\ud q_1\, q_1^2\left(\int\tilde\ud q_2\int\tilde\ud q_3\,\fii(q_{23})+\int\tilde\ud q_2\int\tilde\ud q_3\int\tilde\ud q_4\,\fii(q_{23})\fii(q_{34})+\dots \right)-\\
2\int\tilde\ud q_1\int\tilde\ud q_2\,\fii(q_{12})q_1q_2\left(\int\tilde\ud q_3 + \int\tilde\ud q_3\int\tilde\ud q_4\,\fii(q_{34})+\dots \right)-\\
2\int\tilde\ud q_1\int\tilde\ud q_2\int\tilde\ud q_3\,\fii(q_{12})\fii(q_{23})q_1q_3\left(\int\tilde\ud q_4 + \int\tilde\ud q_4\int\tilde\ud q_5\,\fii(q_{45})+\dots \right)+\dots\,,
\end{multline}
where we reshuffled the series: the first line contains the $d=1$
terms, the second line contains the $d=2$ terms and so on. Let us add
and subtract the term $2\int\tilde\ud q_1\int\tilde\ud q_2\, q_1^2$
from the first line.  Then all the parentheses
become equal to $1/\gamma$ (c.f. Eq.~\eqref{eq:rhonormq}). This can be
factored out leaving another infinite series
\be
-\frac2\gamma \int\tilde\ud q_1\,q_1\left(q_1+\int\tilde\ud q_2\,
\fii(q_{12})q_2 +
\int\tilde\ud q_2
\int\tilde\ud q_3\,
\fii(q_{12})\fii(q_{23})q_3+\dots\right) = 
-\frac2\gamma \int\tilde\ud q_1\,q_1h_1(q_1)\,.
\ee
Thus the contribution of the first term in Eq.~\eqref{eq:2ndterm} is 
\be
-\frac2\gamma \int\tilde\ud q\,q\,h_1(q) + 2\int\tilde\ud q_1\int\tilde\ud q_2\, q_1^2\,.
\label{eq:II/1}
\ee

\item $2q_1q_{d+1}$.\\
We can proceed in the same way for the second term in the first
parenthesis of Eq.~\eqref{eq:2ndterm}.
\begin{multline}
2\int\tilde\ud q_1\, q_1\left(\int\tilde\ud q_2\,q_2\int\tilde\ud q_3\,\fii(q_{23})+\int\tilde\ud q_2\,q_2\int\tilde\ud q_3\int\tilde\ud q_4\,\fii(q_{23})\fii(q_{34})+\dots \right)+\\
2\int\tilde\ud q_1\,q_1\int\tilde\ud q_2\fii(q_{12})\left(\int\tilde\ud q_3\,q_3 + \int\tilde\ud q_3\int\tilde\ud q_4\,q_3\fii(q_{34})+\dots \right)+\\
2\int\tilde\ud q_1\,q_1\int\tilde\ud q_2\int\tilde\ud q_3\,\fii(q_{12})\fii(q_{23})\left(\int\tilde\ud q_4\,q_4 + \int\tilde\ud q_4\,q_4\int\tilde\ud q_5\,\fii(q_{45})+\dots \right)+\dots\,,
\end{multline}
where we reshuffled the series similarly to the previous case. Let us
add and subtract now the term $2\int\tilde\ud q_1\int\tilde\ud q_2\,
q_1q_2$, so that all the parentheses above become equal to
$\int\tilde\ud q\,2\pi\rho(q)q$. After factoring these out we are left
with the same infinite series, so the contribution of the second term
in Eq.~\eqref{eq:2ndterm} is
\be
2\left(\int\tilde\ud q\,2\pi\rho(q)q\right)^2 - 2\int\tilde\ud q_1\int\tilde\ud q_2\, q_1q_2\,.
\label{eq:II/2}
\ee

\end{itemize}

Thus the total contribution of the second line of Eq.~\eqref{eq:FF} is
\be
-\frac2\gamma \int\tilde\ud q\,q\,h_1(q) +
2\left(\int\tilde\ud q\,2\pi\rho(q)q\right)^2 + \int\tilde\ud q_1\int\tilde\ud q_2\, (2q_1^2-2 q_1q_2)\,.
\label{eq:II}
\ee

\subsection{The final result: closed expression for
  \texorpdfstring{$g_3(\gamma,\tau)$}{g3}}

The final result is given by the sum of the partial results
\eqref{eq:I} and \eqref{eq:II}. The last terms of these expressions
exactly cancel each other because
\be
\int\tilde\ud q_1\int\tilde\ud q_2\,q_{12}^2 = \int\tilde\ud q_1\int\tilde\ud q_2\,(2q_1^2-2q_1q_2)\,,
\ee
so we arrive at the dimensionless form of Eq.~\eqref{eq:result}
\be
\frac{g_3(\gamma,\tau)}{\gamma^3} =
 \int\tilde\ud q\,(q^4+q^2)2\pi \rho(q)-
\int\tilde\ud q\, \left(4q^2+1+\frac2\gamma\right)q\,h_1(q)+
3\int\tilde\ud q\, q^2h_2(q)+2\left(\int\tilde\ud q\,2\pi\rho(q)q\right)^2\,.
\label{eq:resultq}
\ee

\section{Galilean invariance}

In this section we show that our expressions \eqref{eq:g2genq},
\eqref{eq:resultq} are invariant under a Galilean boost, which provides
a non-trivial consistency check of our results.

The integral equations after a boost with momentum $b$ read as
\be
\tilde h_m(q)=q^m + \int\frac{\ud q'}{2\pi}\,f(q'-b)\fii(q-q')\tilde h_m(q')\,.
\ee
Note that as a special case $h_0(q)=2\pi\rho(q)$.
After a simultaneous shift in both variables $q$ and $q'$ we
obtain
\be
\tilde h_m(q+b)=(q+b)^m + \int\frac{\ud q'}{2\pi}\,f(q')\fii(q-q')\tilde h_m(q'+b)\,.
\ee
From the iterative solution of these equations it is easy to see that
\begin{subequations}
\begin{align}
\tilde h_0(q+b) &= h_0(q)\,,\\
\tilde h_1(q+b) &= h_1(q)+b\,h_0(q)\,,\\
\tilde h_2(q+b) &= h_2(q)+2b\,h_1(q)+b^2\,h_0(q)\,.
\end{align}
\label{eq:hboost}
\end{subequations}

Now we can calculate the boosted version of $g_2$. From Eq.~\eqref{eq:g2genq}
\be
\frac{\tilde g_2}{\gamma^2} = \int\frac{\ud q}{2\pi}\,f(q-b)
\left[2\pi\tilde\rho(q)\,q^2 -
f(q-b)\tilde h_1(q)\,q\right]\,.
\ee
Shifting the integration variables and using Eqs.~\eqref{eq:hboost} we
arrive at
\be
\frac{\tilde g_2}{\gamma^2} = \frac{g_2}{\gamma^2} + 
\int\frac{\ud q}{2\pi}\,f(q)\left[h_0(q)(2bq+b^2) - 
h_1(q)\,b-h_0(q)(q+b)b\right]\,,
\ee
where we used $h_0(q)=2\pi\rho(q)$. Now due to $\int\ud
q\,f(q)h_0(q)\,q = \int\ud q\,f(q)h_1(q)$ the integral can be readily
shown to vanish, implying the Galilean invariance of $g_2$.

The invariance of $g_3$ can be shown along the same lines.

\section{Local correlations in the ``caricature state''}

In this section we discuss the derivation of the asymptotic values of
the local correlations for an infinite momentum kick. Let us denote
the various functions in this excited state by the superscript
``ex''. The dimensionless integral equations can be written as
\be
h^{\text{ex}}_m(q)=q^m + \left(\int_{-q_2}^{-q_1}\frac{\ud
  q'}{2\pi}+\int_{q_1}^{q_2}\frac{\ud q'}{2\pi}+\right)
\fii(q-q') h^{\text{ex}}_m(q) = q^m + \int_{q_1}^{q_2}\frac{\ud 
  q'}{2\pi} \,\left[\fii(q-q') + (-1)^m \fii(q+q')\right]h^{\text{ex}}_m(q')\,.
\ee
In particular, $h^{\text{ex}}_0(q)=2\pi\rho^{\text{ex}}(q)$ with normalization
\be
\int_{q_1}^{q_2}\frac{\ud q}{2\pi}\,h^{\text{ex}}_0(q)=\frac1{2\gamma}\,.
\ee
We want to use as a reference state the ground state solution on the
interval $[-(q_2-q_1)/2,(q_2-q_1)/2]\equiv[-d,d]$, so we change our variables as
\be
q=k+\frac{q_1+q_2}2\equiv k+a\,.
\ee
This leads to 
\begin{align}
h^{\text{ex}}_m(k+a)\equiv h_m(k) =& (k+a)^m + \int_{-d}^{d}\frac{\ud 
  k'}{2\pi}\, \left[\fii(k-k') + (-1)^m \fii(2a+k+k')\right]h_m(k')=\nonumber\\
&(k+a)^m + \int_{-d}^{d}\frac{\ud k'}{2\pi}\, 
\left[\fii(k-k') + \frac{(-1)^m}{2a^2} 
\left(1-\frac{k+k'}a+\frac{3(k+k')^2-1}{4a^2}\right)\right]h_m(k')
\end{align}
and $\int_{-d}^{d}\frac{\ud k}{2\pi}\,h_0(k)=1/(2\gamma)$.

Expanding the resolvent of the integral equation in $a^{-1}$ we obtain
the solution
\be
h_m(k)=h_m^{\text{b}}(k)+\frac{(-1)^m}{2a^2}(1-\hat\fii)^{-1}\circ
\int_{-d}^{d}\frac{\ud k'}{2\pi}
\left(1-\frac{k+k'}a+\frac{3(k+k')^2-1}{4a^2}\right)h_m^{\text{b}}(k')
+\frac1{4a^4\gamma_0}\int_{-d}^{d}\frac{\ud k}{2\pi}\,h_m^{\text{b}}(k)+\mathcal{O}(a^{-5})\,,
\label{eq:hmsol}
\ee
where $h^{\text{b}}_m(k)$ is the solution of the equation obtained by
keeping only non-negative powers of $a$:
\be
h^{\text{b}}_m(k)=(1-\hat\fii)^{-1}\circ (k+a)^m\,. 
\ee
These are nothing else but the Galilean boosted functions discussed in
the previous section (c.f. Eq.~\eqref{eq:hboost}):
\be
h_0^{\text{b}}(k)=h^{(0)}_0(k)\,,\quad
h_1^{\text{b}}(k)=h^{(0)}_1(k)+ah^{(0)}_0(k)\,,\quad
h_2^{\text{b}}(k)=h^{(0)}_2(k)+2ah^{(0)}_1(k)+a^2h^{(0)}_0(k)\,,
\label{eq:hb}
\ee
where the superscript denotes the ground state solutions on the
interval $[-d,d]$.
The last term in Eq.~\eqref{eq:hmsol} is the result of multiple
applications of the operator $(1-\hat\fii)^{-1}$, and we introduced the notations
\be
\int_{-d}^{d}\frac{\ud k}{2\pi}h^{(0)}_0(k)=\frac1{\gamma_0}\,,\quad
\int_{-d}^{d}\frac{\ud k}{2\pi}h^{(0)}_1(k)k=e_1\,,\quad
\int_{-d}^{d}\frac{\ud k}{2\pi}h^{(0)}_0(k)k^2=e_2\,.
\ee
Using Eqs.~\eqref{eq:hb} for $h^{\text{b}}_m(k)$ we find from
Eq.~\eqref{eq:hmsol}
\begin{subequations}
\begin{align}
h_0(k)&=h_0^{\text{b}}(k)+\frac1{2\gamma_0a^2}h_0^{(0)}(k)-\frac1{2\gamma_0\,a^3}h_1^{(0)}(k)+
\frac1{8a^4}\left[\left(3e_2-\frac1{\gamma_0}+\frac2{\gamma_0^2}\right)h_0^{(0)}(k)+
\frac3{\gamma_0}h_2^{(0)}(k)\right]+\mathcal{O}(a^{-5})\,,\\
h_1(k)&=h_1^{\text{b}}(k)-\frac1{2\gamma_0a}h_0^{(0)}(k)+\frac1{2\gamma_0a^2}h_1^{(0)}(k)-
\frac1{a^3}\left[\left(\frac38e_2-\frac12e_1-\frac1{8\gamma_0}-\frac1{4\gamma_0^2}\right)h_0^{(0)}(k)+\frac3{8\gamma_0}h_2^{(0)}(k)\right]+\mathcal{O}(a^{-4})\,,\\
h_2(k)&=h_2^{\text{b}}(k)+\frac1{2\gamma_0}h_0^{(0)}(k)-\frac1{2\gamma_0a}h_1^{(0)}(k)+
\frac1{a^2}\left[\left(\frac78e_2-e_1-\frac1{8\gamma_0}+\frac1{4\gamma_0^2}\right)h_0^{(0)}(k)+\frac3{8\gamma_0}h_2^{(0)}(k)\right]+\mathcal{O}(a^{-3})\,.
\end{align}
\end{subequations}
Comparing the integal of $h_0^{(0)}(k)$ and $h_0(k)$ we can relate
$\gamma$ and $\gamma_0$:
\be
\gamma_0=2\gamma+\frac1{2a^2}+\mathcal{O}(a^{-3})\,.
\ee 

Now we are in the position to calculate the local correlations. For
$g^{\text{ex}}_2(\gamma,a\to\infty)$ we only need the leading
corrections to $h_0(k)$ and $h_1(k)$ and we obtain
\be
\frac{g^{\text{ex}}_2(\gamma,a)}{2\gamma^2}=
2\int_{q_1}^{q_2}\frac{\ud q}{2\pi}\left[h^{\text{ex}}_0(q)q^2-
h^{\text{ex}}_1(q)q\right] =
2\int_{-d}^{d}\frac{\ud k}{2\pi}\left[h_0(k)(k+a)^2-h_1(k)(k+a)\right]
=
\frac{g_2(\gamma_0)}{2\gamma_0^2}+\frac1{2\gamma_0^2}+\mathcal{O}(a^{-1})\,,
\ee
which gives $g^{\text{ex}}_2(\gamma,a\to\infty)=g_2(2\gamma)/2+1$. For
$g^{\text{ex}}_3(\gamma,a\to\infty)$ we need all the corrections 
given above. The divergent positive powers of $a$ cancel,
and after a lengthy calculation we arrive at the remarkably simple result
\be
g^{\text{ex}}_3(\gamma,a\to\infty)=\frac14 g_3(2\gamma)+\frac94 g_2(2\gamma)\,.
\ee
It is interesting to note that as $\gamma\to0$, the asymptotic values
are given by
$g^{\text{ex}}_3(\gamma\to0,\infty)\to3/2$ and
$g^{\text{ex}}_3(\gamma\to0,\infty)\to5/2$. On the other hand, in the
free boson case we recover $g^{\text{ex}}_2(0,\infty)=2/2+1=2$ and
$g^{\text{ex}}_2(0,\infty)=1/4\cdot6+9/4\cdot2=6$.

\end{document}